\documentclass[10pt,twocolumn]{article} 
\usepackage[a4paper, total={6.5in, 9in}]{geometry}

\usepackage[switch]{lineno}
\usepackage{graphicx}
\usepackage{bbold} % for identity matrix
\usepackage{bm} % bold in equations
\newcommand{\vect}[1]{\boldsymbol{\mathbf{#1}}}
\usepackage{chemformula} % chemical formla
\usepackage{booktabs}
\usepackage{makecell}
\usepackage{textcomp, gensymb}
\usepackage{amsmath}
\usepackage{gensymb} % degree symbol
\usepackage[affil-it]{authblk}
\usepackage{url,hyperref}
\usepackage{lipsum}
\usepackage{cite}
\usepackage{xcolor}
\usepackage{lmodern}
\usepackage{titlesec}
\hypersetup{
	colorlinks,
	linkcolor={black},
	citecolor={blue!40!black},
	urlcolor={blue!60!black}
}

\usepackage{parskip}% http://ctan.org/pkg/parskip
\usepackage[font={small, sf}, labelfont={small,bf}, labelsep=space]{caption} %Chaning caption font
\graphicspath{{figures/}}
\title{Accurate Energy Barriers for Catalytic Reaction Pathways: An Automatic Training Protocol for Machine Learning Force Fields}

\author[1,*]{Lars L Schaaf}
\author[2]{Edvin Fako}
\author[2,*]{Sandip De}
\author[2]{Ansgar Schäfer}
\author[1]{Gábor Csányi}

\affil[1]{%
    Engineering Laboratory, University of Cambridge, Cambridge, CB2 1PZ UK}
\affil[2]{%
    BASF SE, Carl-Bosch-Straße 38, 67056 Ludwigshafen, Germany}
\affil[*]{corresponding authors: lls34@cam.ac.uk (L. L. Schaaf), sandip.de@basf.com (S. De)}

\usepackage{xr}
\makeatletter

\newcommand*{\addFileDependency}[1]{% argument=file name and extension
\typeout{(#1)}% latexmk will find this if $recorder=0
\@addtofilelist{#1}
\IfFileExists{#1}{}{\typeout{No file #1.}}
}\makeatother

\providecommand{\keywords}[1]
{
  \small	
  \textbf{\textit{Keywords:  }} #1
}

\begin{document}
% \linenumbers

\twocolumn[
\begin{@twocolumnfalse}
	\maketitle
	\begin{abstract}

        In this study, we introduce a training protocol for developing machine learning force fields~(MLFFs), capable of accurately determining energy barriers in catalytic reaction pathways. The protocol is validated on the extensively explored hydrogenation of carbon dioxide to methanol over indium oxide. With the help of active learning, the final force field obtains energy barriers within 0.05~eV of Density Functional Theory. Thanks to the computational speedup, not only do we reduce the cost of routine in-silico catalytic tasks, but also find a 40\% reduction in the previously established rate-limiting step. Furthermore, we illustrate the importance of finite-temperature effects and compute free energy barriers. The transferability of the protocol is demonstrated on the experimentally relevant, yet unexplored, top-layer reduced indium oxide surface. The ability of MLFFs to enhance our understanding of extensively studied catalysts underscores the need for fast and accurate alternatives to direct \textit{ab-intio} simulations.\\
        \\
		\vspace{1cm}
	\end{abstract}

\end{@twocolumnfalse}
]

% \tableofcontents
\section{Introduction}
Computational modeling plays a central role in understanding heterogeneous catalysis at an atomic scale. By complementing experimental observations, simulations are used to discover detailed reaction mechanisms, rationalize catalytic trends, and 
guide the design of catalytic materials~\cite{pinheiroaraujoFlameSprayPyrolysis2022, dangRationallyDesignedIndium2020}. 
Limited by the computational cost of \textit{ab-intio} methods, however, catalytic systems are often represented by small idealized surfaces with individual molecules adsorbed at a specific active site. Under these approximations, Density Functional Theory (DFT) is extensively used to draw connections between adsorption properties and catalytic activity~\cite{hammerImprovedAdsorptionEnergetics1999, chenComputationalMethodsHeterogeneous2021, norskovOriginOverpotentialOxygen2004}. However, neglecting more complex effects, such as interactions between different adsorbates, poisoning, finite temperature, and complex surface morphology, can severely limit the relevance of computational studies to experimental observations~\cite{chenComputationalMethodsHeterogeneous2021,zhuoEffectCOCoverage2013, bonatiNonlinearTemperatureDependence2023, reuterCompositionStructureStability2001b, timmermannIrOSurfaceComplexions2020, stockerEstimatingFreeEnergy2023}. Furthermore, selecting viable reaction paths and specific intermediate configurations requires expert knowledge or extensive trial and error. Finding all relevant reaction mechanisms and accurately predicting reaction rates would require extensive screening~\cite{johnsonPyntaAutomatedWorkflow2023}. For large reaction networks and complex surfaces, this goes beyond DFT capabilities.  

An alternative is to use empirically parameterized force fields, which are orders of magnitude faster. However, while there exist reactive force fields~\cite{senftleReaxFFReactiveForcefield2016}, their parameters need to be adjusted for every novel system and often deviate significantly from the true PES due to their limited expressivity~\cite{vandermauseActiveLearningReactive2022b}. 

Machine learning force fields (MLFFs) offer a way to bridge this gap. Rather than starting a new electronic structure calculation for each step in a simulation, the MLFF predicts energy and forces for novel configurations using a model trained on a set of reference configurations.  Apart from being orders of magnitude faster for small systems, local MLFFs scale linearly with system size, are reactive, and systematically improvable by augmenting the training set.  
Recent innovations in describing atomic environments and training frameworks have allowed MLFF to describe more and more complex interactions in both molecules and materials~\cite{behlerGeneralizedNeuralNetworkRepresentation2007, bartokGaussianApproximationPotentials2010, drautzAtomicClusterExpansion2019, batznerEquivariantGraphNeural2022b,batatiaMACEHigherOrder2022b, deringerGaussianProcessRegression2021, gelzinyteTransferableMachineLearning2023}. In the context of heterogeneous catalysis, MLFFs have evolved from capturing low dimensional cuts of the PES to the direct simulation at the micron-scale~\cite{vandermauseActiveLearningReactive2022b, batznerEquivariantGraphNeural2022b, lorenzRepresentingHighdimensionalPotentialenergy2004}. The pioneering neural network potential by \textit{Lorenz et al.} described the six degrees of freedom of the hydrogen molecule on a fixed catalyst surface~\cite{lorenzRepresentingHighdimensionalPotentialenergy2004}. This early-stage MLFF was able to describe hydrogen dissociation in this simplified system. More recently \textit{Vandermause et al.} ran direct reactive molecular dynamics (MD) of 80 $H_2$ molecules over an interacting platinum surface, replicating experimental dissociation, recombination, and diffusion rates~\cite{vandermauseActiveLearningReactive2022b}. In addition to MLFFs trained for specific catalytic reactions, large training sets with up to 55 different elements have emerged~\cite{tranMethodsComparingUncertainty2020b, mamunHighthroughputCalculationsCatalytic2019,chanussotOpenCatalyst20202021,tranOpenCatalyst20222022}. Accurate potentials trained on such data-sets can be used for screening of desired catalytic properties. Nevertheless, the accuracy in adsorption energy and energy barriers currently remains significantly lower than can be obtained for system-tailored MLFFs~\cite{chanussotOpenCatalyst20202021, vandermauseActiveLearningReactive2022b}.
%To date the application to heterogeneous catalysis has been limited to single reaction steps, small molecules or transition metal surfaces~\cite{vandermauseActiveLearningReactive2022b, guAutomatedExploitationBig2022}. 

In this paper, we focus on training an MLFF for a specific system. We go beyond describing a single reaction step and focus on exploring the entire reaction pathway of CO\textsubscript{2} hydrogenation to methanol with a single potential. 
We develop a hands-off training protocol for rapidly generating MLFFs, that can predict adsorption energies and energy barriers using nudged elastic bands (NEBs), starting only from a set of suggested intermediates along the desired reaction path. The intermediate geometries could be obtained from a force field geometry optimization or manual placement. We show the efficacy of active learning\cite{bernsteinNovoExplorationSelfguided2019, zhangActiveLearningUniformly2019a,smithLessMoreSampling2018} in converging the modeled potential energy surface to that of DFT in the relevant domains. Ultimately, the final force field's minimum energy paths (MEP) correspond to MEPs on the true DFT potential energy surface. The corresponding energy barriers are converged within one $kT$ (45~meV) at reaction conditions (500K), where the entire reaction path spans 50~$kT$.

We validate our training protocol using a well-studied reaction mechanism, namely the carbon dioxide hydrogenation to methanol on an indium oxide surface~\cite{yeDFTStudyCO2012,yeActiveOxygenVacancy2013, zhangInsightDFTInitio2018, freiMechanismMicrokineticsMethanol2018, douDFTStudyCatalyzed2018, dangRationallyDesignedIndium2020, caoRelationsSurfaceOxygen2021}. 
In the context of reducing carbon dioxide emissions and carbon recycling, methanol synthesis from waste carbon oxide (CO and CO\textsubscript{2}) streams is of significant industrial importance. Indium oxide has emerged as a promising catalyst, particularly for operation under CO\textsubscript{2}-rich conditions, offering improved selectivity compared to copper-based catalysts that currently dominate industrial practice~\cite{nakamuraIssueActiveSite2003, studtMechanismCOCO22015}. This reaction pathway, comprising ten intermediates and five distinct reactions, presents significant challenges for training a single MLFF due to the diversity of the adsorbate species, and the complexity of the surface structure. 
As a benchmark, we investigate the oxygen vacancy on cubic-In2O3(110) with the lowest formation energy, as reported by Dang et al.~\cite{dangRationallyDesignedIndium2020}. Additionally, we showcase the transferability of the protocol on a previously unexplored surface. Recent experimental and computational investigations have revealed that, under experimental conditions, the top surface of indium oxide undergoes complete reduction~\cite{caoRelationsSurfaceOxygen2021a, bielzHydrogenReducibilityBonding2010}. By applying our protocol to this reduced surface, we demonstrate its capability to describe a more realistic and experimentally relevant surface. 

We illustrate the utility of a fast, yet accurate MLFF, by exploring multiple minimum energy paths for each reaction and calculating free energy barriers. This systematic investigation uncovers notable alterations in the reaction pathway. Specifically, we identify a preferred minimum energy path for the rate-limiting step, characterized by a substantially lower barrier. Additionally, investigations at finite temperatures, reveal that the true rate-limiting step corresponds to the production of formaldehyde. 

\section{Results}

\subsection{Active Learning}

	\label{sec:active learning}
	\begin{figure}
		\centering
		\includegraphics[width=3.2in]{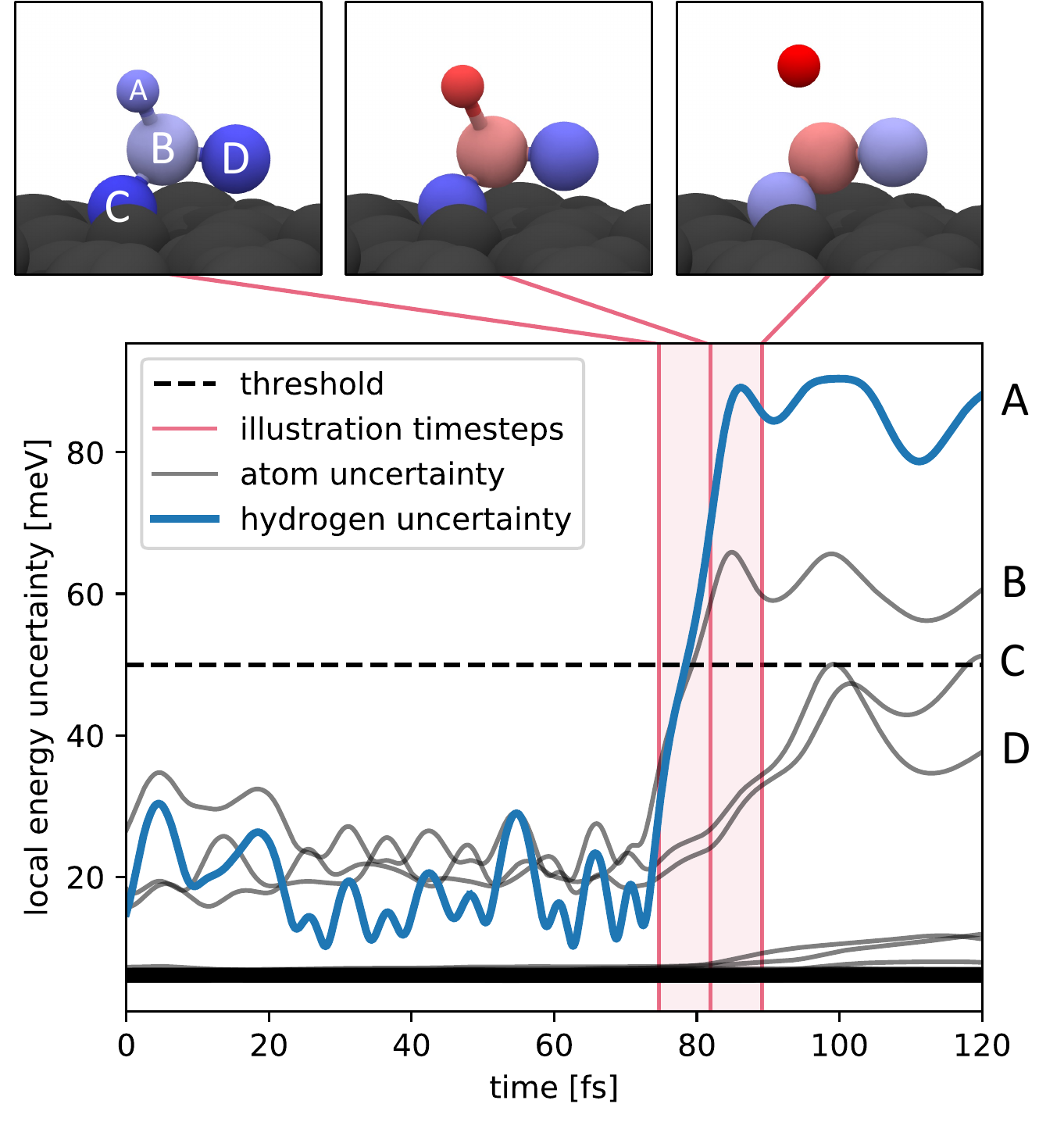}
		\caption{\textbf{Predicted energy uncertainty during molecular dynamics (MD).}  This figure demonstrates the initial iteration of active learning with molecules, highlighting the occurrence of a nonphysical bond break when the energy uncertainty surpasses a threshold of 50~meV. The top panel displays MD snapshots of formate (HCOO) on In\textsubscript{2}O\textsubscript{3}, where molecular atoms are color-coded by the local energy uncertainty, while the surface atoms are depicted in grey. 
  } 
		\label{fig:active learning-ill}
	\end{figure}

It is common for ML force fields to be trained in an iterative manner, where novel configurations are sampled from MLFF simulations and added back into the training set. The relevant geometries can be selected manually or based on explicit criteria, such as changes in density or bonding topology~\cite{bartokMachineLearningGeneralPurpose2018, deringerMachineLearningBased2017, deringerGeneralpurposeMachinelearningForce2020,deringerOriginsStructuralElectronic2021, behlerFirstPrinciplesNeural2017, schuttSchNetDeepLearning2018, chmielaSGDMLConstructingAccurate2019, bernsteinNovoExplorationSelfguided2019}. While successful, this approach relies on human expert knowledge, which is inherently limited to a low-dimensional understanding of force field failures. More recently, active learning has been used to automate this process using a model's uncertainty metric~\cite{botuMachineLearningForce2017, podryabinkinAcceleratingCrystalStructure2019, jorgensenExplorationExploitationGlobal2018, vandermauseOntheflyActiveLearning2020, sivaramanMachinelearnedInteratomicPotentials2020,youngTransferableActivelearningStrategy2021, vandermauseActiveLearningReactive2022b, vanderoordHyperactiveLearningHAL2022}. 

    We employ a sampling strategy based on the local energy uncertainty of individual atoms~\cite{jinnouchiOntheflyMachineLearning2019, vandermauseActiveLearningReactive2022b,smithLessMoreSampling2018}. When the uncertainty of any atom exceeds a predefined threshold, new configurations are sampled from the ongoing simulation. These new geometries are subsequently evaluated with Density Functional Theory (DFT) and incorporated into the training set. In cases where the threshold is not reached during the simulation, we sample either the configuration with the highest uncertainty (for molecular dynamics simulations) or the final configuration (for optimization tasks). The machine learning force field (MLFF) is then retrained using the expanded dataset, and the aforementioned steps are repeated. This active learning loop continues until a desired level of accuracy is achieved.

    Our specific active learning approach, as depicted in Figure~\ref{fig:final-path-gap-vs-dft}b, incorporates two distinct stopping criteria to guide the automatic iterative training of the MLFF. The first criterion, referred to as the uncertainty threshold ($\sigma_{thr}$), determines when to interrupt a simulation and sample a configuration. Throughout the training protocol, the uncertainty threshold is set to 50 meV, a choice that is justified in Figure~\ref{fig:active learning-ill}. The second criterion, referred to as the termination criteria, determines when to conclude the active learning loop and considers the MLFF to be sufficiently accurate for the intended task. The termination criteria are tailored to the specific simulation type (see Section~\ref{sec:protocol}) and can be determined based on the observed DFT-MLFF error in the sampled configurations. The exact termination criteria depend on the user's desired level of accuracy.

    In this manner, the model's accuracy at the desired task converges to the specified accuracy, without prior knowledge of the reaction path energetics. Active learning reduces the accumulation of uninformative training data, while systematically improving the model's accuracy in relevant parts of the potential energy surface. The method section contains more detail on how the underlying sparse Gaussian process provides a cheap metric for the model's predicted error.

     The effectiveness of using local energy uncertainty to determine when to interrupt simulations is demonstrated through the initial iteration of running molecular dynamics (MD) with adsorbates. Figure~\ref{fig:active learning-ill} depicts the evolution of energy uncertainty for each atom during the MD. Notably, the four molecular atoms exhibit significantly higher uncertainty compared to the 79 bulk atoms. Throughout the MD trajectory, the local uncertainties remain relatively low, until the energy uncertainty of the hydrogen atom exceeds the predetermined threshold of 50~meV at 78~fs.

    The substantial increase in uncertainty for the hydrogen atom suggests that the simulation has encountered an atomic environment that greatly differs from any seen in the training set. This indicates the need to sample a new configuration for the next iteration of the potential. It is noteworthy that only a few atoms exhibit a pronounced increase in uncertainty, highlighting the importance of monitoring individual atom energy errors rather than total errors. The specific threshold value becomes less critical due to the sharp rise in uncertainty observed. In this particular system, a threshold of 50~meV ensures the sampled configurations remain chemically relevant and informative for further training iterations.

\subsection{Training Protocol}

 \begin{figure*}[t]
	\centering
    \includegraphics[width=.85\linewidth]{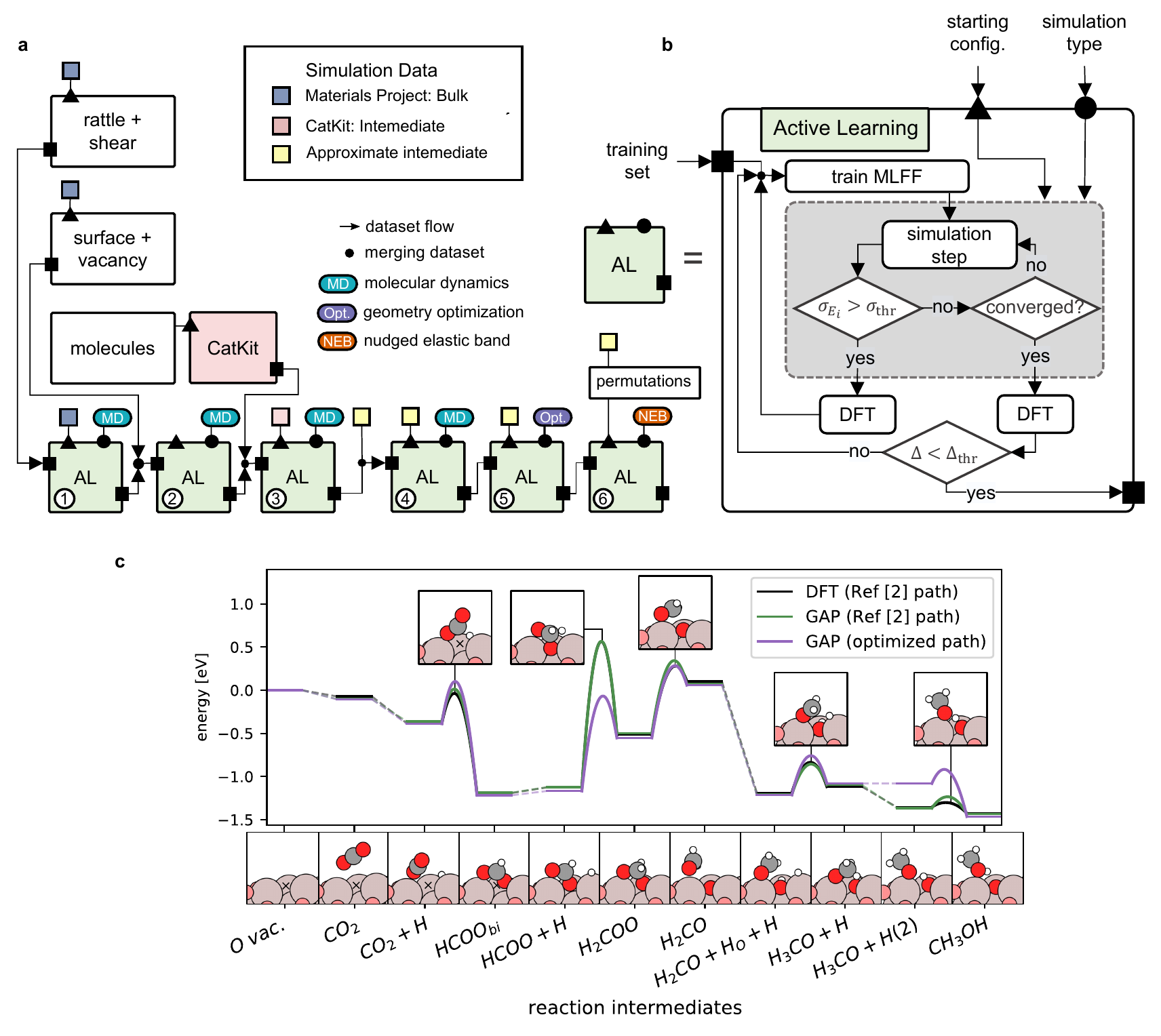}
	\caption{\textbf{Training protocol.} A flow chart (\textbf{a}) of the five active learning~(AL) blocks that make up the protocol. Each active learning block (\textbf{b}) iteratively samples new training configurations, when the model uncertainty surpasses a threshold ($\sigma_{thr}$) until a sampled configuration satisfies the termination criteria ($\Delta_{thr}$). Simulations include running MD, geometry optimization, and nudged elastic bands. The protocol is validated, by replicating the Dang et al.~\cite{dangRationallyDesignedIndium2020} path (green) and used to optimize the path (purple), finding a 40\% reduction in the rate-limiting energy barrier (\textbf{c}). The location of the oxygen vacancy is marked with a cross and the opacity of the surface atoms is reduced.}
	\label{fig:final-path-gap-vs-dft}
\end{figure*}
    
	\label{sec:protocol}
    The training protocol consists of five active learning blocks which sample configurations from MD, geometry optimization, or nudge elastic band calculations. The flow chart in Figure~\ref{fig:final-path-gap-vs-dft}a shows how these blocks link in series to obtain the final training set. The only inputs to the protocol are the bulk configuration of indium oxide and the set of approximate geometries corresponding to the intermediates along the reaction path.  These intermediate geometries can come from force field minimization, manual placement, or adsorption prediction tools like CatKit.
    
    Throughout the protocol, we only ever perform single-point calculations with DFT, while all dynamical simulations are performed with the MLFF.   
    Below we outline the protocol's five active learning blocks and their termination criteria.
    Once a sampled configuration satisfies the termination criteria, the training protocol moves on to the next block. While we use very strict termination criteria, these will depend on the user's desired accuracy. 
    
    The first two blocks model the surface itself, while the remaining three are used to capture inter-molecular and molecule-surface interactions. The first three blocks consist of running active learning with MD to gain a stable potential and the last two correspond to geometry optimization and nudge elastic band calculations to obtain accurate adsorption energies and barriers. We find that running active learning with MD before geometry optimization helps sample repulsive behavior effectively. The number of configurations in the training set, at each stage of the protocol, is tabulated in Supplementary Note~1.

	\paragraph{0. Initial training set} The initial training set consists of the bulk, conventional unit cell of cubic In\textsubscript{2}O\textsubscript{3}, taken from the Materials Project~\cite{jainCommentaryMaterialsProject2013}. Additionally, we included ten configurations obtained by  rattling all atom positions with random displacements drawn from a Gaussian with standard deviation $0.02$~\AA{} and applying random deformation to the unit cell, with a standard deviation of $0.01$. This ensures we sample repulsive behavior immediately.  Additionally, we obtain a set of initial geometries for the intermediates, referred to as approximate intermediates, which enter the protocol in blocks~3-5.

	\paragraph{1-2. Bulk and surface dynamics} In the first active learning block, we iteratively increase our training set until the potential can run stable dynamics. 
    The potential is deemed stable if, during a 10~ps molecular dynamics (MD) run at 300K, the uncertainty threshold is never breached for all intermediates. We repeat the same procedure for the slab and oxygen vacancy of interest, requiring one additional iteration, using the same termination criteria. 
	
	\paragraph{3. Adsorbate dynamics} We now extend our model to capture molecule-surface interaction, creating a four-element force field~(H, C, O, In). To reduce the number of iterations needed to obtain stable MD we enhance the training set before active learning with:
    \begin{itemize}
        \item Dimers: configurations consisting of two atoms spaced such that their repulsive forces never exceed 20~eV\AA{}\textsuperscript{-1}. We include five such dimer configurations for each possible combination between carbon, oxygen, and hydrogen, totaling 24 configurations. We set the energy and force weights of all dimer configurations at a factor of five lower than the rest of the training set. 
        \item Catkit configurations: obtained by placing the molecules from the intermediates around the active site using Catkit's adsorbate placement tool~\cite{boesGraphTheoryApproach2019a}. Rather than placing molecules directly on the relaxed surface, we use the potential from the second block, to run MD on the active site first. This avoids sampling unnecessarily similar atomic environments from the surface.  After placing the intermediates within 3\AA{} of the oxygen vacancy with Catkit, we select three per intermediate using farthest point sampling of all atomic environments. Here we use the distances between SOAP vectors of the local environments to measure dissimilarity, following the same procedure as De et al.~\cite{deComparingMoleculesSolids2016}. 
    \end{itemize}
	
    We now run MD with the MLFF on all approximate intermediates for 1~ps. As previously we terminate active learning if, during MD, the predicted uncertainty stays below the threshold for all intermediates. As mentioned in Section~\ref{sec:active learning}, using the uncertainty estimate is crucial for sampling new configurations in this block. 
 
	\paragraph{4. Adsorbate Geometry optimization} Using geometry optimization, we now try to find the DFT adsorbate structure starting from the approximate intermediates. During active learning, we monitor the performance of the potential by evaluating the GAP minima configurations with single-point DFT. If the DFT forces are sufficiently small for all intermediates, the MLFF's minima correspond to true minima in the DFT potential energy surface and we terminate the active learning loop. However, if the DFT forces are above 0.2~eV\AA{}\textsuperscript{-1} for any intermediate, we add the configurations to the training set and continue the active learning loop. Note that the adsorbate structures further improve during the last active learning block. 
	
	 \paragraph{5. Minimum Energy Path Search} Before starting active learning, we add a set of linearly interpolated images between the reaction intermediates to the training set. Running active learning, we then perform NEB calculations, sample the resulting minimum energy paths, and add them to the training set. The active learning loop is terminated when the total energy error (DFT vs. MLFF) on all sampled configurations is less than 50~meV. Note that for each reaction, we run NEBs for all possible permutations of like atoms of the molecule~(see section~\ref{sec:lower-rate-limiting} for more detail). During sampling we only DFT evaluate the NEB path with the lowest energy.

\subsection{Validation on known pathway}
\label{sec:valid-dang}
We validate the protocol, on an extensively explored reaction pathway: \ch{CO2} hydrogenation to methanol on indium oxide with a a single oxygen vacancy~\cite{yeActiveOxygenVacancy2013, freiMechanismMicrokineticsMethanol2018, dangRationallyDesignedIndium2020, caoRelationsSurfaceOxygen2021}. 
Figure~\ref{fig:final-path-gap-vs-dft}, shows the close alignment between DFT (black) and the final MLFF (green) throughout the entire reaction path. The final training set consists of 622 configurations. Supplementary Note~1 outlines the number of training configurations needed at each stage of the protocol. 
 	
\paragraph{Adsorption Energies}
The adsorption energies converge throughout the training protocol. Adsorbate-surface interactions are first sampled in active learning (AL) blocks three and four, after which the mean adsorption energy error is 82~meV. We improve upon this with AL and geometry optimization in block five. To allow for a direct comparison, relaxations are initialized from configurations similar to Reference~\cite{dangRationallyDesignedIndium2020} (see Supplementary Note~2 for further details on starting configurations). 
After eleven iterations we reach the termination criterion outlined in Section~\ref{sec:protocol}, at which point the mean adsorption energy error is 23~meV across intermediates. These errors further reduce to 12~meV throughout NEB active learning. See Supplementary Figures 1 and 2 for more details on the adsorption energy convergence across iterations.
 	
\paragraph{Minimum Energy Path}
\begin{figure}
	\centering
	\includegraphics[width=0.87\linewidth]{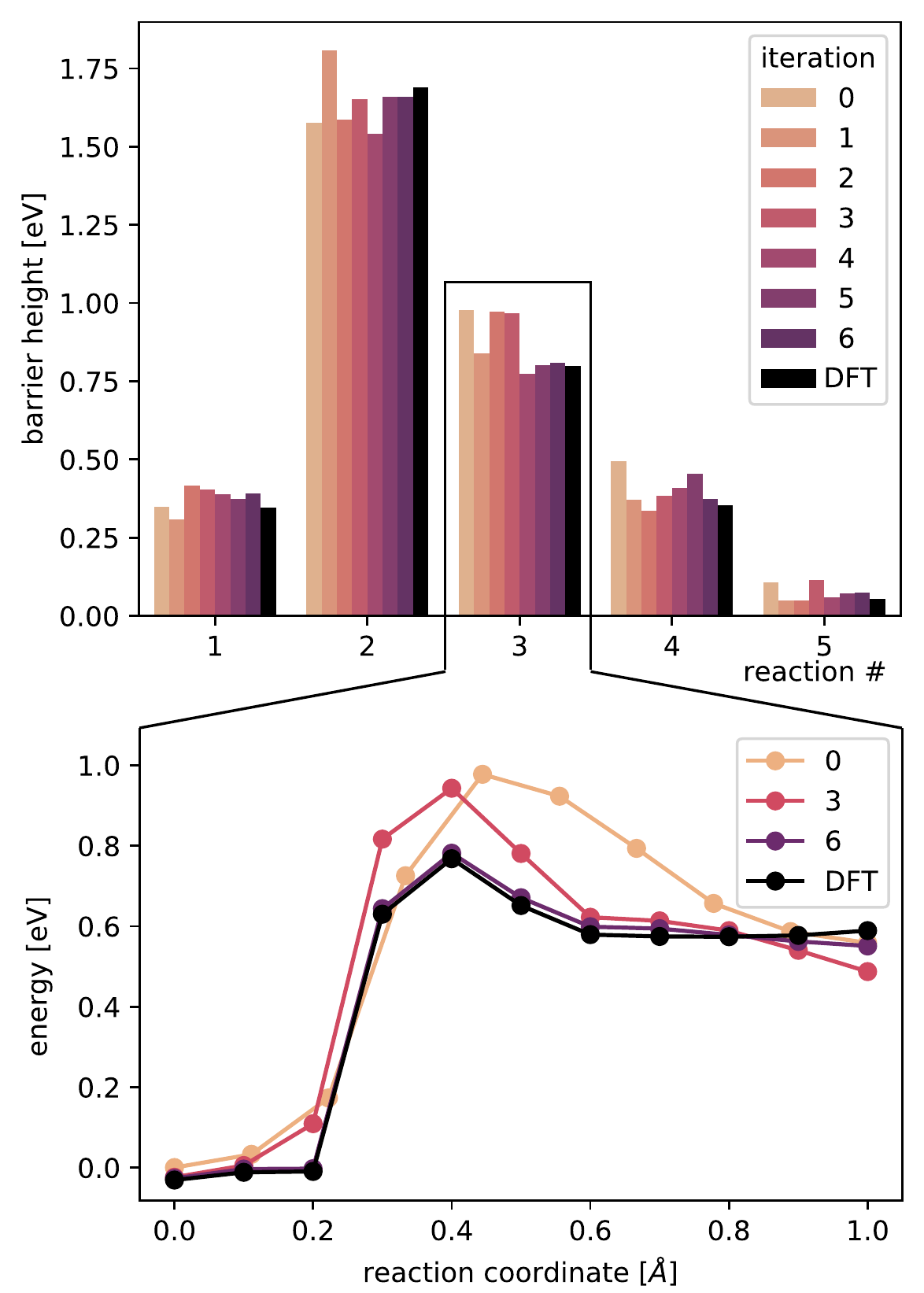}
	\caption{\textbf{Energy barrier accuracy throughout active learning.} Barrier height comparison between DFT and GAP models at different AL iterations for all energy barriers (top panel). The bottom panel gives a detailed view of the different minimum energy paths throughout active learning for the third reaction (\ch[arrow-offset=5pt]{H\textsubscript{2}COO -> H\textsubscript{2}CO + O}).}
	\label{fig:iteration-neb-progression}
\end{figure}

Throughout the protocol, the energy barriers converge to those obtained with DFT. To validate the final force field, we initialize DFT-NEB calculations with the MLFF's minimum energy paths (MEPs). We find that the DFT-NEBs are converged without a single optimization step and have a projected force below 0.05~eV\AA{}\textsuperscript{-1} for all five reactions. The MLFF has hence found true MEPs on the DFT potential energy surface. 
Note that as we aim to validate the protocol against Reference~\cite{dangRationallyDesignedIndium2020} in this section, we don't permute all atoms with similar species, as detailed in the protocol, but choose a permutation that is similar to the previously published literature.

Figure~\ref{fig:iteration-neb-progression}, shows the convergence of the MEP throughout active learning iterations. We find that without any NEB active learning, it is possible to draw qualitative conclusions, with a mean barrier error of 50~meV. At this stage of the protocol, the training set only contains 325 configurations. After six iterations of NEB active learning, the GAP MEPs converge to those of DFT, providing an energy barrier estimate within 45~meV for all five reactions. As highlighted in Section~\ref{sec:comp-cost}, training an MLFF following this automated protocol requires approximately eight times fewer DFT single-point calculations, than running DFT-NEBs directly.

\subsection{Lower rate-limiting step}
\label{sec:lower-rate-limiting}
We initiate the training protocol afresh, assuming the adsorbate configurations are unknown for this reaction. To generate initial guesses for the reaction intermediates, we employ CatKit and select the adsorbate configuration closest to the position of the oxygen vacancy. After the successful termination of the protocol, the acquired training set contains 696 configurations. Figure~\ref{fig:final-path-gap-vs-dft}c compares the reaction pathway of the resulting MLFF (purple) to the pathway explored by Dange et al.~\cite{dangRationallyDesignedIndium2020} (green). The paths align closely except for a significant reduction in the rate-limiting barrier and the last reaction path. The difference in the last reaction step originates from a subtle rearrangement of the \ch{H3CO + H} intermediate for the path found in the literature. As we start from cat-kit configurations, in the training protocol these two intermediates are identical. Below we discuss in greater detail how the automatic protocol finds an optimized reaction path for the rate-limiting step. 

\begin{figure}
	\centering
	\includegraphics[width=1\linewidth]{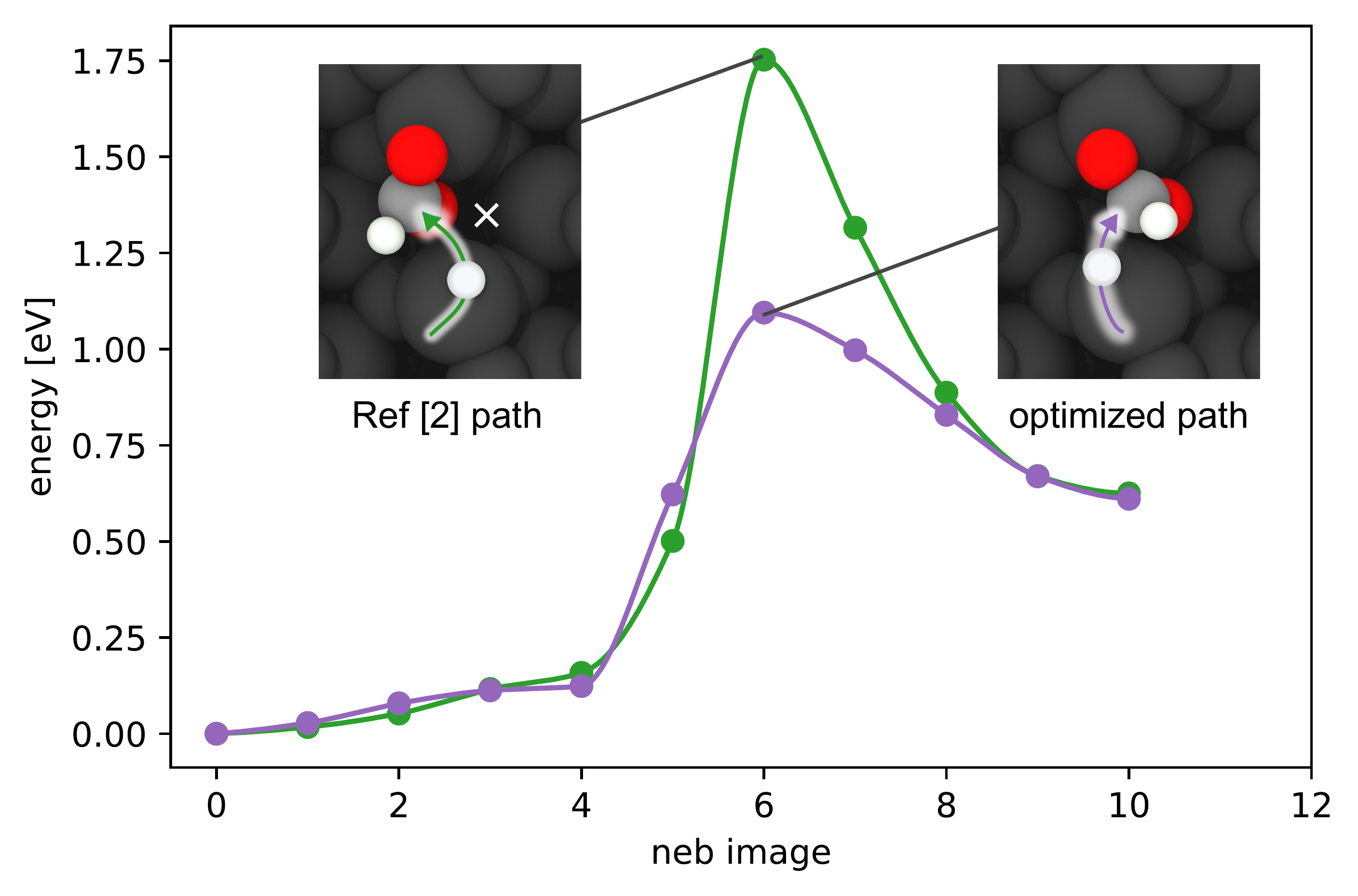}
	\caption{\textbf{Lower rate-limiting step.} Comparison of two NEBs for the rate-limiting step, with different permutations of like atoms. The protocol investigates all possible permutations, finding a significantly lower barrier (green) than Reference~\cite{dangRationallyDesignedIndium2020} (blue). Visualizations of the transition states (insets), show the the hydrogen atom's trajectory and the location of the oxygen vacancy (white cross).}
	\label{fig:lower-rls}
\end{figure}

When performing NEB calculations in an automatic fashion, a crucial consideration arises regarding the ordering of like atoms. Since the atoms in the initial image are connected to atoms in the final image along a smooth path, any alteration in the index of atoms in the initial image leads to a distinct reaction path. Traditionally, the selection of the most plausible arrangement has relied on chemical intuition. However, given the computational efficiency of NEB calculations with the Machine Learning Force Field (MLFF), we adopt a comprehensive approach by performing all possible permutations of like atoms of the molecule. Subsequently, we identify the path that exhibits the lowest energy as the preferred path. 

This comprehensive approach unveils a preferential reaction path for the rate-limiting step (\ch[arrow-offset=5pt]{HCOO + H -> H2COO}). This optimized path, depicted in purple in Figure~\ref{fig:lower-rls}, has a significantly reduced barrier of 1.0~eV. The insets reveal a key difference between the two paths: throughout the optimized path, one of the intermediate's oxygen atoms occupies the oxygen vacancy. By screening permutations, we achieve a systematic approach to investigate barriers, avoiding the potential pitfalls of possibly misleading chemical intuition. We note that for larger intermediates, screening all possible permutations may become unfeasible. Here we suggest pre-selecting the most relevant permutations by comparing the geometric distance between the initial and final NEB configurations. 

\subsection{Finite temperature effects}
\label{sec:finite-temp}

Until now we have assumed that the zero kelvin potential energy surface can be used to infer properties of the catalyst. This approximation is often made due to the computational cost of DFT. We now illustrate the importance of incorporating finite temperature effects by investigating the free energy barrier of dioxymethylene to formaldehyde conversion (\ch[arrow-offset=5pt]{H\textsubscript{2}COO -> H\textsubscript{2}CO + O}). In doing so, we find that the energy barrier is larger than previously assumed and that formaldehyde production is the rate-limiting step. 

\begin{figure}
	\centering
	\includegraphics[width=1\linewidth]{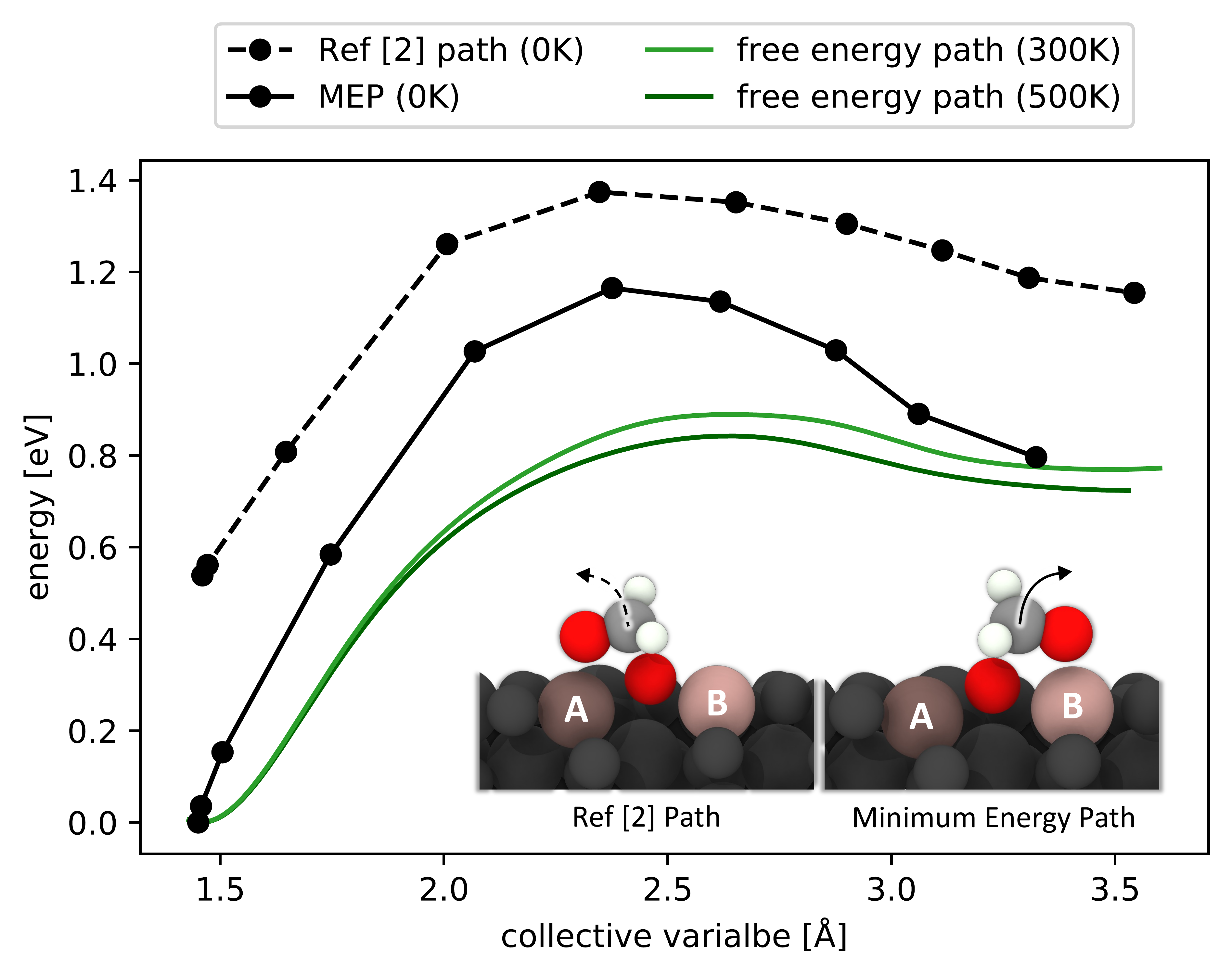}
	\caption{\textbf{Free energy barriers for formaldehyde production.} Comparison of the Reference~\cite{dangRationallyDesignedIndium2020} path (dotted) and the minimum energy path (MEP, bold). The black lines represent NEB energies, while the green lines depict the free energy barriers along the MEP, calculated by umbrella integration. The insets display the transition states of Ref.~\cite{dangRationallyDesignedIndium2020} and the MEP, highlighting the distinct oxygen-coordination of the two indium atoms (atom A is four-fold coordinated, atom B is three-fold coordinated).}
	\label{fig:finite-temp}
\end{figure}

We perform umbrella sampling at ambient and operating (500K) temperatures, using the distance between the oxygen atom and the carbon atom as a collective variable. 
Within the first picosecond of molecular dynamics, the starting configuration, taken as the local minima found in literature, moves to a lower energy state with different geometry. 
The insets in Figure~\ref{fig:finite-temp} illustrate that the H\textsubscript{2}COO intermediate preferentially binds to the indium atom, with lower oxygen coordination. Consequently, the minimum energy path (depicted by the bold line) differs from the path investigated by Reference~\cite{dangRationallyDesignedIndium2020} (represented by the dotted path). With a barrier height of 1.1~eV, the production of formaldehyde (\ch{H2CO}) rather than \ch{H2COO} is rate-limiting. For a more detailed visual comparison of the paths, please refer to Supplementary Figures 3-7. This exploration of the potential energy surface, facilitated by the computational speed of MLFF, reveals that although there may exist some high-energy local minima, their relevance to the overall reaction progression is limited. Finite temperature simulations promptly reveal the thermodynamically relevant intermediate geometries.  

By running umbrella sampling, we circumvent the approximation that reaction rates are solely determined by a single transition state and local energy minima on the potential energy surface. The temperature dependence of the free energy barrier can have a substantial impact on the predicted selectivity and activity of catalysts, as well as the behavior of micro-kinetic models employed to analyze the progression of reactions~\cite{liSumFrequencyGeneration2021, khatamiradDatadrivenHighthroughputWorkflow2023,xieAchievingTheoryExperiment2022, kozuchHowConceptualizeCatalytic2011}.

\subsection{Transfer-ability to unexplored surfaces}

To demonstrate the transferability of the automatic training protocol, we apply it to a previously unexplored surface. Additionally, we illustrate that the obtained training set can be effectively utilized by different machine learning methods, indicating its versatility beyond the specific GAP framework. 

\paragraph{Reduced surface}
\begin{figure}
    \centering
    \includegraphics[width=1\linewidth]{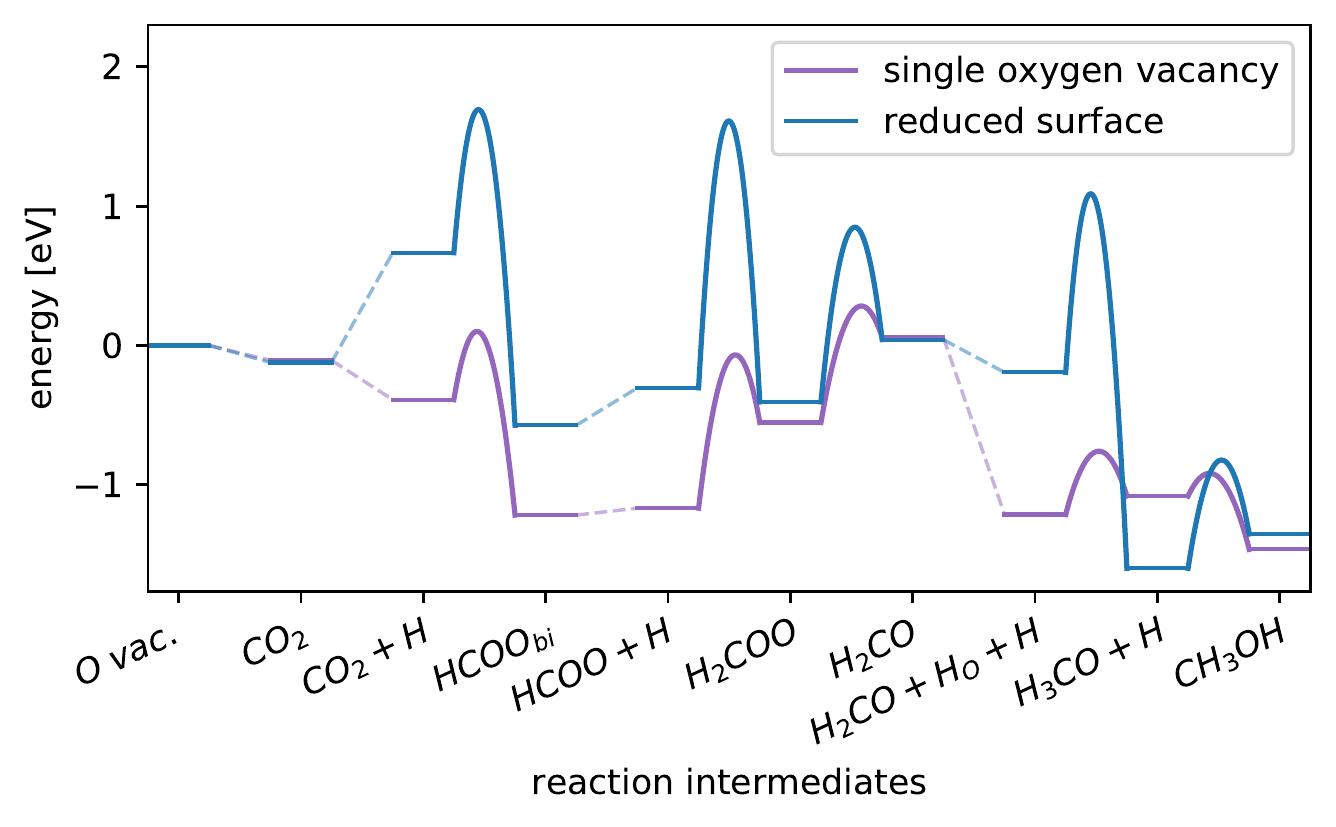}
    \caption{\textbf{Reduced surface reaction pathway.} Comparing the reaction pathway for the single oxygen vacancy and the top-layer reduced surface. }
    \label{fig:results-reduced} 
\end{figure}

Recent computational investigations have revealed the thermodynamic preference for the reduced surface under experimental conditions~\cite{caoRelationsSurfaceOxygen2021}. To evaluate the transferability of our protocol, we apply it to the top-layer reduced surface, yielding the reaction pathway depicted in Figure~\ref{fig:results-reduced}. The construction of the training set for this force field required less than one thousand single-point DFT calculations, with 70\% of the calculations dedicated to the final active learning phase to accurately determine energy barriers. Notably, the pathway reveals a large barrier for the hydrogenation of carbon dioxide.

\paragraph{Doped surfaces}

\begin{figure}
	\centering
	\includegraphics[width=1\linewidth]{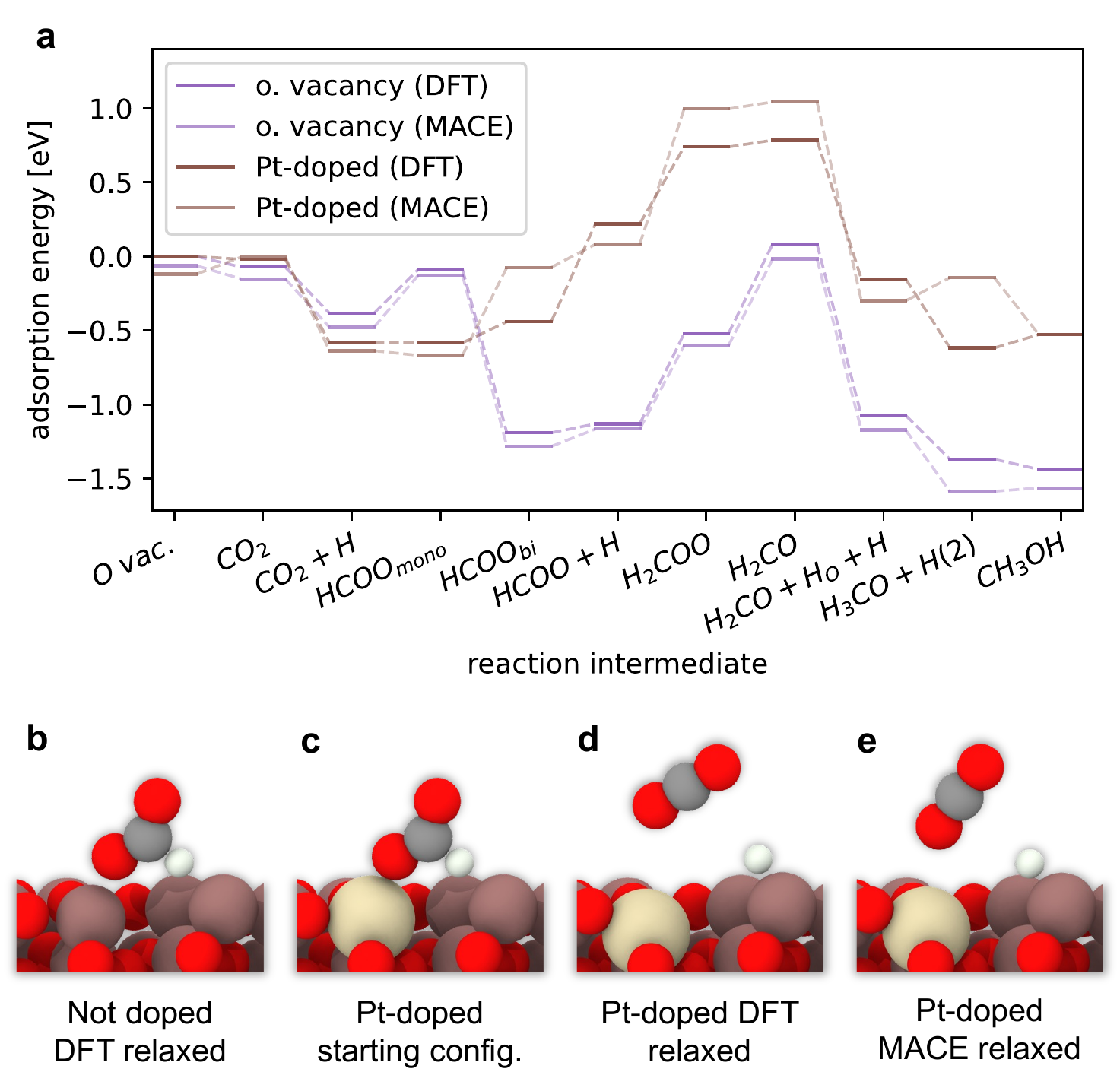}
	\caption{\textbf{Training set transferability to other MLFF frameworks} A comparison between the MACE~\cite{batatiaMACEHigherOrder2022b} model and DFT for both undoped and doped surfaces (\textbf{a}). The intermediates on the Pt-doped surface differ significantly, such as formate splitting into CO\textsubscript{2}~+~H~(\textbf{b-e}).}
	\label{fig:doped-graph-configurations}
\end{figure}

We examine the transferability of our training dataset to other machine learning force field frameworks. Specifically, we use MACE, a  higher-order equivariant message-passing neural network, to train an MLFF on the dataset acquired in Section~\ref{sec:valid-dang} for the single oxygen vacancy surface. Without any additional training data, it achieves a mean adsorption energy error of 38 meV.

Furthermore, we investigate the behavior of doped surfaces. Dopants play a crucial role in modifying the adsorbate structure and activation of intermediates, leading to significant changes in catalytic activity and selectivity~\cite{hannaganFirstprinciplesDesignSingleatom2021}. In the case of indium oxide, platinum, and palladium dopants have been demonstrated to enhance catalytic performance by promoting H\textsubscript{2} splitting~\cite{freiNanostructureNickelpromotedIndium2021, freiAtomicscaleEngineeringIndium2019}. These dopants not only increase the abundance of adsorbed hydrogen, needed for CO\textsubscript{2} hydrogenation and vacancy formation but also influence the entire reaction pathway.

To avoid repeating the entire training protocol, which would be data-inefficient, we explore the possibility of enhancing the training set without additional active learning steps. We find that with sixty additional configurations, the MACE framework captures general trends in the adsorption energies, as visible in Figure~\ref{fig:doped-graph-configurations}. The additional training configurations originate from the oxygen vacancy training set, where the Indium atom closest to the active site is replaced by a platinum atom and randomly rattled with a standard deviation of 1\AA{}. It is worth noting that improving the adsorption energies through active learning with the MACE model could be pursued, but is beyond the scope of this paper. Application to a different MLFF framework demonstrates the transferability of the acquired training data.

\subsection{Exploration of adsorbate structures}
\begin{figure}
	\centering
	\includegraphics[width=0.97\linewidth]{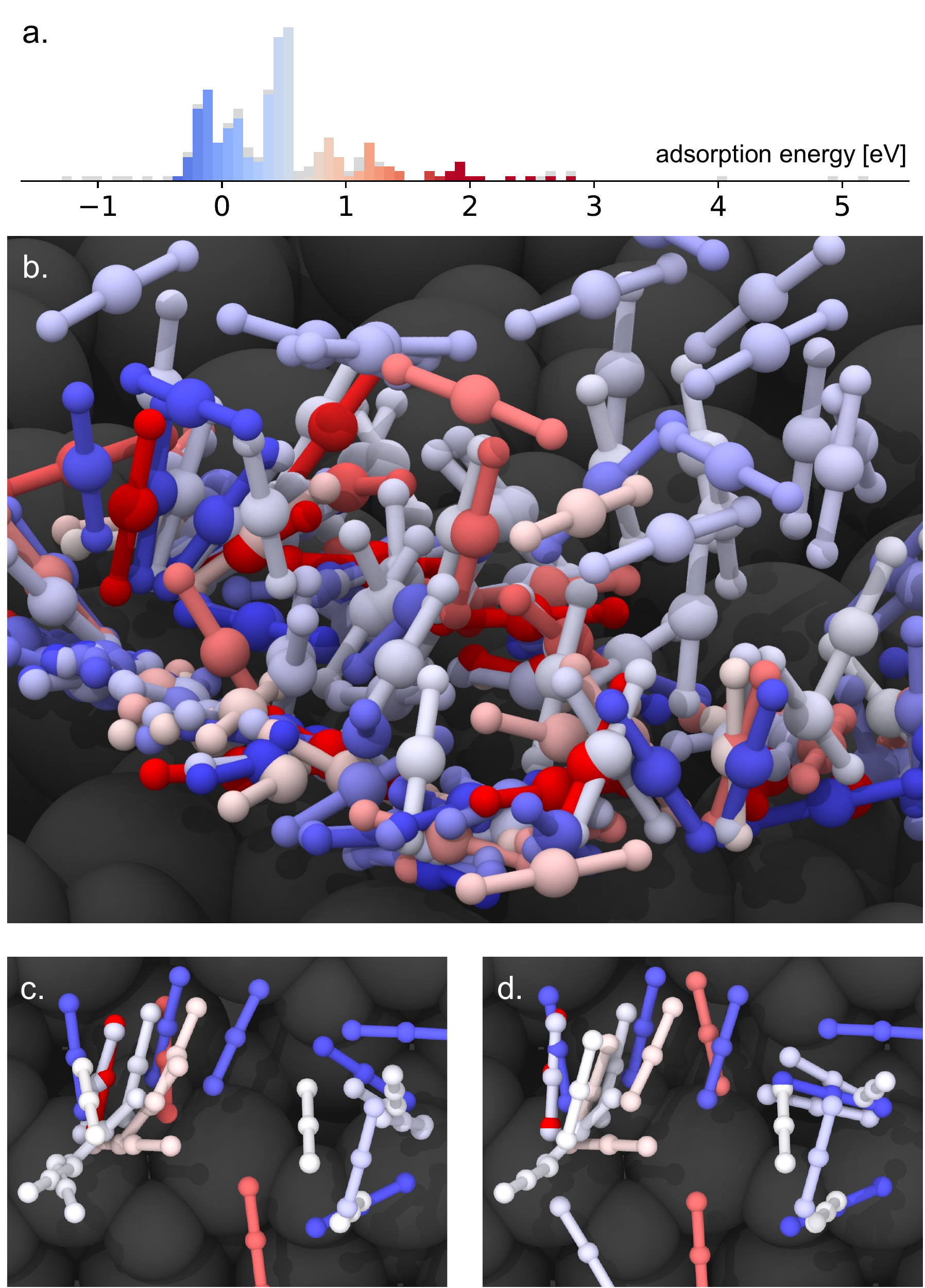}
	\caption{\textbf{Adsorbate configuration exploration.} Displaying 236 unique MLFF adsorption geometries (\textbf{b}) of carbon dioxide found with minima hopping  and their adsorption energy distribution (\textbf{a}). Figures (\textbf{b-c}) contain only the molecular atoms of the adsorbate states, color-coded by adsorption energies, and the relaxed oxygen vacancy in dark grey as a backdrop. All carbonate configurations are excluded, and their energies are shaded in grey in the histogram \textbf{(a)}.  A subset of MLFF minima (\textbf{c}) is relaxed with DFT (\textbf{d}) for validation. }
	\label{fig:mh-both}
\end{figure}

We showed in Section~\ref{fig:finite-temp} that thermodynamically relevant adsorption configurations can elude DFT investigations, as adsorbates may get stuck in local energy minima. 
To investigate whether this is a more general property of these complex surfaces, we use the MLFF to sample adsorption configurations of the very first intermediate: \ch{CO2}. A thorough sampling of the surface reveals a plethora of local minima as illustrated in Figure~\ref{fig:mh-both}a.
Each such minimum is a potential starting structure for studying the subsequent reaction step.  This suggests that running a small number of geometry optimization with expensive ab-initio methods is unsuited for conclusively finding adsorbate configurations.

We systematically screen local minima using minima hopping, which consists of alternating high-temperature MD with geometry optimization~\cite{goedeckerMinimaHoppingEfficient2004, jungMachinelearningDrivenGlobal2023}. 
We use CatKit to sample ten starting configurations for each intermediate and then run ten iterations of active learning with minima hopping. 
The final potential finds 236 unique minima. All non-unique minima are removed, by comparing the local environment of the carbon atoms using SOAP vectors (see Supplementary Method~2 for details). To validate the new adsorbate geometries, we relax 25 MLFF-found minima with DFT to a maximum force of 0.01~eV\AA{}\textsuperscript{-1}. We find that the MLFF (Figure~\ref{fig:mh-both}c) and DFT minima (Figure~\ref{fig:mh-both}d) are extremely similar. Note that the top two layers of the surface are not fixed and hence differ for all adsorbate configurations. 

Figure~\ref{fig:mh-both} shows all 236 unique adsorption geometries placed on top of each other.  For visualization purposes, only the CO\textsubscript{2} molecules are included and color-coded by adsorption energy, while the geometry-optimized oxygen vacancy in dark grey acts as a backdrop.  
The lowest energy configurations are found when CO\textsubscript{2} reacts with a surface oxygen atom to form carbonate.  Previous studies, that also identified the thermodynamic stability of carbonate, found that it doesn't partake in the conversion to methanol~\cite{dangRationallyDesignedIndium2020}. Carbonate configurations are hence excluded from the figure. As visible from the histogram, a significant number of local minima have negative adsorption energies.

\subsection{Computational cost comparison}
\label{sec:comp-cost}
The findings summarised above rely heavily on the computational speedup gained by the MLFF. To compare the computational cost, we calculate the first energy barrier, starting from a simple interpolation of images with DFT. After 38 steps with BFGS, requiring 1092 single-point DFT evaluations, the NEB converges to 0.08~eV\AA{}\textsuperscript{-1}. As a comparison, the final GAP model requires only 622 DFT training configurations and finds NEBs for all five reaction steps that converged to below 0.05~eV\AA{}\textsuperscript{-1}, when evaluated with single point DFT. Running all DFT-NEBs would hence require approximately five thousand single-point evaluations, eight times more than the size of the training set. 

Beyond its computational efficiency for routine catalytic tasks, the MLFF offers extensive exploration capabilities of the potential energy surface. Considering all feasible atom permutations within the same chemical species helps us obtain a 40\% reduction in the activation energy of the rate-limiting step. Furthermore, determining the free energy barriers required 6.4 million samples of the potential energy surface. For an 84-atom configuration, the GAP model runs 9.2 steps per second on the dual AMD EPYC™ 7742 64-core processors at 2.25GHz. In contrast, a single DFT calculation requires an average of 33 minutes on the same hardware.

\section{Discussion}
We present an automatic training protocol for machine learning force fields (MLFF) capable of accurately capturing the energetics of a given heterogeneous reaction path. We validate our approach on the extensively explored conversion of CO2 to methanol on indium oxide with a single oxygen vacancy. The minimum energy paths (MEPs) for all five reactions found using the final MLFF correspond to true MEPs on the DFT potential energy surface. A single model is hence able to accurately describe a reaction with ten intermediaries on a complex oxide surface. Additionally, we showcase the transferability of the protocol by investigating the same mechanism on a new and previously unexplored surface: top-layer reduced indium oxide. 

We show that MLFFs offer more than just computational cost reduction for routine computational catalysis tasks; they emerge as an essential tool for in-depth mechanistic catalytic investigations. By running multiple nudged elastic band calculations for each reaction, we find a preferred MEP for the rate-limiting step, with a 40\% lower energy barrier. Moreover, through finite temperature sampling, we unveil lower minima for the intermediates of the subsequent largest barrier, identifying it as the true rate-limiting step. Finally, we illustrate the power of MLFFs by computing free energy barriers at ambient and operating temperatures, requiring over six million samples of the potential energy surface. %, using an MLFF that is trained on under seven hundred DFT single-point calculations. 

Accurately describing the true minimum energy path of individual reactions significantly influences the relevance of computational studies to experiment. Predictions from micro-kinetic or energetic span models can depend strongly on specific barriers, resulting in notable differences in observables, such as the turnover frequency~\cite{kozuchHowConceptualizeCatalytic2011, reuterInitioThermodynamicsFirstPrinciples2016}. We anticipate that active learning protocols will facilitate the adoption of MLFFs in the wider catalysis community, enabling more comprehensive mechanistic explorations of catalytic cycles.

\section{Methods}

\subsection{Machine learning framework}

We use the Gaussian Approximation Potential (GAP) framework~\cite{bartokGaussianApproximationPotentials2010} to fit DFT energies and forces obtained using Quantum Espresso~\cite{giannozziAdvancedCapabilitiesMaterials2017}. GAP is chosen due to its maturity, past success at describing a wide variety of systems, and because it provides an analytical uncertainty estimate~\cite{deringerAtomisticUnderstandingDisordered2018, bartokMachineLearningGeneralPurpose2018, monserratLiquidWaterContains2020}. As a many-body descriptor, we use the smooth overlap of atomic positions (SOAP). To assess the transferability of the training set, we employ MACE~\cite{batatiaMACEHigherOrder2022b}, a higher-order equivariant message-passing neural network that leverages recent advancements in graph neural networks for atomic simulations~\cite{schuttSchNetContinuousfilterConvolutional2017b, batznerEquivariantGraphNeural2022b, batatiaMACEHigherOrder2022b}. 

\subsection{Uncertainty estimation}
\label{sec:gap-error}
Various approaches exist to predict a model's error for active-learning~\cite{vanderoordHyperactiveLearningHAL2022, vandermauseActiveLearningReactive2022b, schranCommitteeNeuralNetwork2020, novikovMLIPPackageMoment2021}. The Gaussian Approximation Potential framework allows for a rigorous estimate of the local energy uncertainty from the underlying Gaussian Process Regression (GPR).  
In the GAP framework, the total energy of a configuration with $N$ atoms is given by the sum of atomic energies,
\begin{linenomath*}
\begin{equation}
    E = \sum_i^N \epsilon(\vect{\rho}_i), 
\end{equation}
\end{linenomath*}
where $\vect{\rho}_i$ is a descriptor of an atom's local atomic environment, that only depends on the positions and elements of its neighbors within a specified distance cutoff $r_{cut}$. Thanks to this locality approximation the computational cost of the force field scales linearly with the number of atoms. 

To describe the many-body atomic environment $\vect{\rho}$ we use the smooth overlap of atomic positions (SOAP) descriptor, which is invariant with respect to rotations, permutations, translations, and reflections~\cite{bartokRepresentingChemicalEnvironments2013}.

The local energy function $\epsilon$ is modeled using sparse GPR. 
To quantify the similarity between the two environments, we use a polynomial kernel,
\begin{linenomath*}
\begin{equation}
    k(\vect{\rho}_n, \vect{\rho}_m) = (\vect{\rho}_n \cdot \vect{\rho}_m)^\zeta,
\end{equation}
\end{linenomath*}
where $\zeta$ is the polynomial degree and $\vect{\rho}$ are SOAP vectors. For a set of $N$ and $M$ local environments, we can define the kernel matrix $[\vect{K_{NM}}]_{nm} \equiv k(\vect{\rho}_n, \vect{\rho}_m)$~\cite{deringerGaussianProcessRegression2021}. If one were to perform GPR conditioned on a training set of $N$ local environments $\vect{\rho}_n$ with ground truth energies $\vect{y}$, the predicted energy distribution for unseen atomic environments $\vect{\rho}$ would be Gaussian 
\begin{linenomath*}
\begin{equation} \label{eq:gpr-distribution}
    \epsilon(\vect{\rho}) \sim \mathcal{N}(\mu(\vect{\rho}),\Sigma^2(\vect{\rho})),
\end{equation}
\end{linenomath*}
with mean and covariance,
\begin{linenomath*}
\begin{equation}\label{eq:gpr-mean-var}
    \centering
    \begin{split}
        \mu(\vect{\rho}) &= \vect{k}_N^T [\vect{K}_{N N} + \sigma_n^{-2} \vect{\mathbb{1}} ]^{-1} \vect{y} \\
        \Sigma^2(\vect{\rho}) &= k - \vect{k}_N^T [\vect{K}_{N N} + \sigma_n^{-2} \vect{\mathbb{1}} ]^{-1} \vect{k}_N,\\
    \end{split}
\end{equation}
\end{linenomath*}
where $k$ and $\vect{k}_N$ are the kernels between the environment $\vect{\rho}$, itself and the training set $\vect{\rho}_n$ respectively. While the time complexity of obtaining the true mean $\mu$ and covariance $\Sigma$ functions would scale cubically with training set size, they can be approximated with a sparse method. Rather than quantifying the similarity of novel environments with all $N$ training data points, we use a sparse set of $M$ environments, where $M\ll N$. Modifications to Equation~\ref{eq:gpr-mean-var} can be found, by minimizing the Kullback-Leibler divergence between the sparse method and the full posterior in Equation~\ref{eq:gpr-distribution}~\cite{titsiasVariationalLearningInducing2009}. The approximate mean and covariance functions are,
\begin{linenomath*}
\begin{equation}\label{eq:sgpr-mean-var}
    \centering
    \begin{split}
        \mu(\vect{\rho}) &= \vect{k}_M^T \vect{C}_{MM} \vect{K}_{MN}  \vect{y} \\
        \Sigma^2(\vect{\rho}) &= k - \vect{k}_M^T \vect{K}_{MM}^{-1} \vect{k}_M + \vect{k}_M^T \vect{C}_{MM} \vect{k}_M,\\
    \end{split}
\end{equation}
\end{linenomath*}
where 
\begin{linenomath*}
\begin{equation}\label{eq:sgpr-mean-var-B}
    \centering
    \begin{split}
    \vect{C}_{MM} =  [\vect{K}_{MM} + \sigma_m^{-2} \vect{K}_{MN} \vect{K}_{MN}^T]^{-1},  \\
    \end{split}
\end{equation}
\end{linenomath*}
as found using variational inference~\cite{titsiasVariationalLearningInducing2009}. 
The training cost of the resulting GAP model scales as $\mathcal{O}(NM^2)$.
During simulation, we use the mean of the sparse posterior distribution, while the variance $\Sigma^2$ provides the local energy error estimate for active learning. As is commonly done within the GAP framework, we select a subset of $M$ representative environments from the training set using CUR matrix decomposition~\cite{deringerGaussianProcessRegression2021}. 

\subsection{Computational details}

\paragraph{DFT} All density functional theory calculations, including single point evaluations, geometry optimizations as well as NEB~\cite{berneClassicalQuantumDynamics1998} calculations were done with QuantumEspresso\cite{giannozziAdvancedCapabilitiesMaterials2017}. We use the Perdew-Becke-Ernzerhof (PBE) functional~\cite{perdewGeneralizedGradientApproximation1996}, core electrons are described by projector augmented-waves (PAW)~\cite{blochlProjectorAugmentedwaveMethod1994,kresseUltrasoftPseudopotentialsProjector1999}, the valence monoelectronic states are expanded as plain waves with a maximum kinetic energy of 884~eV (65 Ry). Core electrons are represented by pseudopotentials~\cite{dalcorsoPseudopotentialsPeriodicTable2014} while the valence shell is represented with 4, 6, 13, and 16 electrons for C, O, In, and Pt. We use a dense reciprocal-space mesh, with a maximum spacing of 0.25~\AA{}$^{-1}$ and Gaussian smearing of 0.1~eV with an SCF convergence criterion of $10^{-7}$~eV, to ensure sufficient convergence of gradients, which is essential for training MLFFs. Reference~\cite{dangRationallyDesignedIndium2020} used the Vienna Ab initio Simulation Package (VASP). A calculation on the reduced energy barrier for the rate-limiting step shows that QuantumEspresso calculations correlate well with those obtained from VASP (see Supplementary Method~1).

\paragraph{GAP} We used the Gaussian Approximation Potential (GAP)~\cite{bartokGaussianApproximationPotentials2010, deringerGaussianProcessRegression2021} in combination with a double turbo SOAPs~\cite{caroOptimizingManybodyAtomic2019} as our many body descriptor, with 4~\AA{} and 6~\AA{} cut-offs. To evaluate the potential we use the quippy python interface~\cite{kermodeF90wrapAutomatedTool2020,csanyiExpressiveProgrammingComputational2007}. A complete list of hyperparameters is included in Supplementary Method~2. 

\paragraph{MACE} We use standard settings, with two message passing layers, 256 channels, and equivariant messages of order L=1. All further settings are tabulated in Supplementary Method~3.

\paragraph{Geometry optimization and NEB calculations}
All slabs are centered between 8\AA{} of vacuum with the lowest two layers (40 atoms) fixed. Relaxations were converged to 0.001~eV\AA{}\textsuperscript{-1} and 0.01~eV\AA{}\textsuperscript{-1} when using the MLFF and DFT respectively unless explicitly stated otherwise. For NEBs calculations, the convergence criteria were set to 0.01~eV\AA{}\textsuperscript{-1} and 0.05~eV\AA{}\textsuperscript{-1} respectively. The criteria were set lower for the MLFF due to its smooth potential energy surface and minimal computational cost. 
All NEBs were calculated using the atomic simulation environment~\cite{larsenAtomicSimulationEnvironment2017} using splines and exponential preconditioning~\cite{makriPreconditioningSchemeMinimum2019}. The initial image guesses were made using the IDPP method~\cite{smidstrupImprovedInitialGuess2014}. We use ten and twenty images per NEBs for the oxygen vacancy and reduced surface respectively. 

\paragraph{Umbrella sampling} The free energy barriers are obtained using the Colvar~\cite{fiorinUsingCollectiveVariables2013} package in combination with LAMMPS~\cite{thompsonLAMMPSFlexibleSimulation2022}. We run 110~ps of molecular dynamics for 32 bins along the collective variable, with a 1~fs time step and a spring constant of 75~eV\AA{}\textsuperscript{-2} for all temperatures. Starting configurations are taken from the relevant NEB path and allowed to equilibrate for 10~ps. Using umbrella integration~\cite{stecherFreeEnergySurface2014}, the samples are mapped to the free energy barrier. 

\paragraph{CatKit} Catkit~\cite{boesGraphTheoryApproach2019a} was used to create initial adsorbate structures to start active learning and to initialize the minima-hopping configurations. Here all surface oxygen atoms were explicitly defined. 

\paragraph{Active site}
We investigate the same active site as Reference~\cite{dangRationallyDesignedIndium2020}. Specifically the oxygen vacancy on the 110 surfaces of cubic indium oxide, with the lowest formation energy. The vacancy configuration itself has 79 atoms per unit cell. 

\section{Data availability}
The datasets generated throughout the active learning protocol and used to train the MLFFs are available in the Zenodo repository, \url{10.5281/zenodo.8268726}.

\section{Code availability}
Both MLFF frameworks used during this investigation, GAP and MACE, are publicly available.

\section{Acknowledgements}
We thank the authors from Reference~\cite{dangRationallyDesignedIndium2020} for sharing their NEB and geometry-optimized configurations to help validate the machine learning force field. 

We are grateful for computational support from the UK national high-performance computing service, ARCHER2, for which access was obtained via the UKCP consortium and funded by EPSRC grant reference EP/P022065/1 and EP/X035891/1. Additionally, LS acknowledges support from the EPSRC Centre for Doctoral Training in Chemical Synthesis with grant reference EP/S024220/1 and corporate funding from BASF SE.  

\section{Competing interests}
Authors LS, EF, AS, and SD declare no financial or non-financial competing interests. Author GC is listed as an inventor on a patent filed by Cambridge Enterprise Ltd. related to SOAP and GAP (PCT/GB2009/001414, filed on 5 June 2009 and published on 23 September 2014).

\section{Author contributions}
Authors GC, AS, and SD jointly conceived the research. Authors LS, SD, and EF did the calculations. Author LS drafted the paper. All authors edited the paper.

\bibliography{PhD_Research_abbrv} 
\bibliographystyle{naturemag}

\end{document}